\documentclass[12pt]{article}

\usepackage{siunitx} 
\usepackage{enumerate} 
\usepackage{pgfplots}
\usepackage{pgfplotstable}
\usepackage{tikz,pgfplots}
\usepackage{amsmath}
\usepackage{wrapfig}
\usepackage{comment}
\usepackage{lipsum}
\usepackage{mathtools}
\usepackage{xcolor}
\usepackage{amssymb}
\usepackage{authblk}

\DeclarePairedDelimiter{\ceil}{\lceil}{\rceil}

\usepackage{geometry}
\pgfplotsset{compat=1.14}

\begin{document}

\title{Information-to-energy trade-offs and the optimal alphabet of polymer replication}
\author[1,2]{Damián G. Hernández}
\affil[1]{Department of Applications of Physics and Biology to Health Sciences, Centro Atómico Bariloche, Argentina}
\affil[2]{Instituto Balseiro, Centro Atómico Bariloche, Argentina}

\date{}

\maketitle  

\begin{abstract} 
    We analyze information transmission in a recently proposed coarse-grained model of polymer replication by framing it as a communication channel between templates and copies. By calculating the mutual information in the steady-state limit of long chains, we recover the accurate-random phase diagram and establish that the information per-monomer depends solely on template specificity within the accurate regime. Crucially, even in the accurate region, small error fractions lead to substantial information loss due to the nonlinear relationship between errors and mutual information. Examining the information-to-energy cost ratio reveals non-monotonic behavior as a function of monomer alphabet size, with an optimum determined primarily by the per-monomer assembly free energy. For DNA's four-base alphabet, we find that the observed effective assembly energy (at least $14\,k_B T$) places the system far from the information-transmission optimum, suggesting that biological replication may prioritize the suppression of spontaneous random assembly over information-to-energy efficiency. We also characterize achievable rate-fidelity trade-offs using Shannon bounds, providing a theoretical framework for evaluating future proofreading mechanisms in ensemble models.
\end{abstract}

\section{Introduction}
	 
    The relevance of information for living systems cannot be understated, from sensing noisy signals from the environment to transmitting information reliably between different places or across generations \cite{bialek2012biophysics}. This is particularly true for copolymerization processes, where the amount of generated information is closely linked to the dissipation and minimum energy cost in the production of persistent copies \cite{andrieux2008nonequilibrium, ouldridge2017fundamental}. 

    Here we aim to quantify the information in a recently published coarse-grained model of polymer replication \cite{genthon2025nonequilibrium} that considers a population or ensemble of copies, which makes it particularly suitable for a stochastic description. This model provides a thermodynamic description of a non-equilibrium process \cite{esposito2011second, andrieux2013information}, leading to different regimes of accurate and/or random polymer copies depending on key parameters of the system: the number of monomer types $m$, template specificity $a$, per-monomer free energy of assembly $\Delta\mu_r$, and free energy provided by the fuel $\Delta\mu_F$. While Genthon et al. analyzed accuracy in terms of error fractions, here we quantify the utility of such copies for message transmission. Also, this approach differentiates from frameworks where equilibrium is assumed to quantify the information in the replication process \cite{arias2012entropy}.

\begin{figure}
    \centering
    \includegraphics[width=0.95\textwidth]{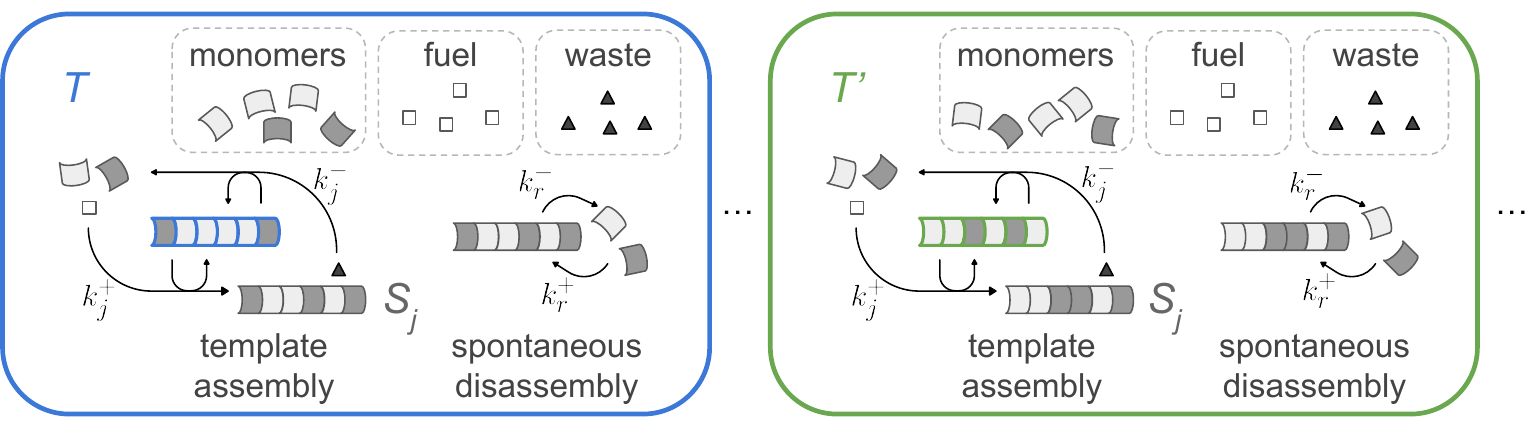}
    \caption{Many instances of the template copying ensemble as an information channel. Templates $T$ are sampled from an initial distribution, and then each template goes through a copying mechanism, producing a population of copies $S_j$. The two pathways involved are template assembly, and spontaneous disassembly. In this scheme, there are two types of monomers.}
    \label{fig0}
\end{figure}   

    In this report, we expand the template copying ensemble by allowing many possible initial templates, each generating its own population of copies. With this setup, we describe the joint distribution of templates and copies and explicitly calculate the mutual information between them in the limit of long chains as a function of the system parameters. Expressing the analysis in terms of information rather than error fractions has two key advantages: (i) it establishes a direct qualitative link to information production and erasure \cite{andrieux2008nonequilibrium}, and (ii) even in the limit of large polymer chains, a small fraction of errors can lead to substantial information loss due to their nonlinear relationship ---a feature that becomes explicit in the information-theoretic framework. For example, a 2\% error rate might be considered accurate, but it can decimate the information capacity by almost 10\%.
    
    Our analysis yields three main results. First, by calculating the partition function in the long-chain limit, we uncover an information-based phase diagram ---that parallels the results from the accurate-random phase diagram of \cite{genthon2025nonequilibrium}--- and establish that the mutual information per-monomer $I/L$ is nonzero only when $\Delta\mu_F > \text{max}(\log m, \Delta\mu_r) -\log[1+e^{-a}(m-1)]$, with the information magnitude depending solely on template specificity $a$ within the accurate regime. Second, we find that the ratio of information to minimum energy cost exhibits non-monotonic behavior as a function of the alphabet size $m$, reaching a maximum at $m^* \sim e^{\Delta\mu_r}$ rather than increasing monotonically. For DNA's four-base alphabet ($m=4$), the actual effective assembly energy is at least $14\,k_B T$, which is much higher than the $1.4\,k_B T$ required to make $m=4$ the information-theoretic optimum. This high value ensures that DNA replication operates in a \emph{quenched} regime where spontaneous random assembly is exponentially suppressed, prioritizing sequence control over energetic efficiency. Third, we characterize the fundamental rate-fidelity trade-offs using Shannon's bound, providing a theoretical framework for evaluating future proofreading mechanisms. Although this behavior emerges from a simplified model and biology may not explicitly optimize information-to-energy ratios, we argue that more realistic copying mechanisms must contend with similar trade-offs regarding alphabet size, specificity, and energetic cost.
        
\section{Joint distribution of templates and copies}
    Conceptually, in order to interpret the model of replication of polymers as an information transmission process, we need to think of many instances of the \emph{template copying ensemble} \cite{genthon2025nonequilibrium} where the original template is chosen at random from a set, and from each template, a population of copies arises (see Fig.~\ref{fig0}). In this way, the channel input is the chosen template, and the output is a randomly sampled copy from the resulting ensemble. How much information does this copy provide us about the template that originated it? Such a procedure allows us to define relevant stochastic variables and their probability distributions. 
    
    First, we denote $T$ as the variable characterizing the possible states of the template $t \in \{1,\dots,m^L\}$, where we consider all possible chains of $m$ monomers of length $L$. To simplify and following \cite{genthon2025nonequilibrium}, we assume that the copies have the same length and exist in the same space, effectively coarse-graining the intermediate steps. Let us denote $S$ as the variable for the copies. Thus, our goal is to calculate the mutual information $I(T;S)$ in the steady state. 

    Briefly recalling, the template copying model provides a coarse-grained, thermodynamically consistent description of polymer replication. A template sequence $T$ of length $L$ resides in a system coupled to reservoirs of monomers ($M$), fuel ($F$), and waste ($W$). The model describes the production of copy sequences $S_j$ through two competing pathways: (i) \emph{template assembly}, a fuel-driven, sequence-dependent pathway where the template acts as a catalyst for the assembly and disassembly of copies; and (ii) \emph{spontaneous disassembly}, a sequence-independent pathway representing background assembly and disassembly without fuel consumption. To maintain thermodynamic consistency, the rates for these processes satisfy local detailed balance: $k_j^+/k_j^-=\exp[-\beta(\Delta\mu_r-\Delta\mu_F)L]$, and $k_r^+/k_r^-=\exp[-\beta\Delta\mu_r L]$, where $\beta^{-1}=k_B T$, $\Delta\mu_r$ is the per-monomer assembly energy and $\Delta\mu_F$ is the chemical potential provided by the fuel. Sequence specificity is introduced via kinetic coefficients $k_j=k_0 e^{-aq}$, where $q$ is the Hamming distance (number of errors) between the template and the copy, and $a$ represents the copying specificity.

    We consider the initial state of the template to be given by a distribution $p(t)$ \footnote{Here, we interpret replication as a channel, such that the input distribution is later chosen to maximize the information between templates and copies \cite{mackay2003information}.}. For a complete description of the joint probability distribution, we need to calculate the conditional distribution of copies given a template $p(s|t)$. As we have mentioned, we consider that we are in the steady state, that is, in the asymptotic time limit. We also consider the large $L$ limit of very long chains and disregard corrections of order $L^{-1}$. From the assembly and spontaneous disassembly processes within the template ensemble model, due to symmetry, the only factor that determines the probability of a copy $s$ given a template $t$ is the Hamming distance $q(s,t)$. Considering that the solution for the number of copies is a Poisson distribution \cite{genthon2025nonequilibrium}, the probability of all $s$ that are at a distance $q=q(s,t)=x(s,t)L$ is proportional to the rate $\lambda_q$,

\begin{equation}
    \displaystyle p(s|t)=\frac{\lambda_{x(s,t)L}}{\sum_{z=0}^L \lambda_z \Omega_z},
    \label{eq1}
\end{equation}
    \noindent such that the number of copies $s$ that are at a distance $q(s,t)$ is $\Omega_q= \binom{L}{q}(m-1)^q$.

    In general, the rate as a function of time follows
\begin{equation}
    \displaystyle \lambda_q(t)= \frac{k_a(q)}{k_d(q)}\left(1-e^{-k_d(q)t} \right),
    \label{eq2}
\end{equation}    
    \noindent where $k_a=k_j^++k_r^+$ and $k_d=k_j^-+k_r^-$ are the total assembly and disassembly rates, respectively. However, under our assumptions (time $t\to\infty$, $L\gg 1$, $k_B T=1$), the rate $\lambda_q$ simplifies to
\begin{equation}
    \displaystyle \lambda_{xL}= e^{-\Delta\mu_r L} \times  
    \begin{cases}
      \frac{k_0}{k_r}e^{\Delta\mu_F L}\,e^{-a x L} & \text{if $x<x_M$}\\
      1 & \text{if $x>x_M$}
    \end{cases}    
    \label{eq3}
\end{equation}    
    \noindent where $x_M=\Delta\mu_F/a$, following the notation used in \cite{genthon2025nonequilibrium}. When $\Delta \mu_F> ax$, we can clearly see that the rate, and thus the conditional probability $p(s|t)$, decreases exponentially with the distance $x(s,t) L$ between the template and the copies.

    Now let us solve the partition function (or normalization) in Eq.~(\ref{eq1}), $\mathcal{Z}=\sum_{z=0}^L \lambda_z \Omega_z$. The main tool for this calculation is the Laplace method, which is used to approximate integrals that have a term inside that grows exponentially with a large parameter ---in this case, the length $L$. Here, the partition function can be expressed as two integrals in $x$,
\begin{equation}
    \displaystyle \mathcal{Z}=\mathcal{Z}_1+\mathcal{Z}_2=\int_0^{x_M} dx\, \lambda_{xL} \Omega_{xL}+\int_{x_M}^1 dx\, \lambda_{xL} \Omega_{xL}.
    \label{eq4}
\end{equation}    
    Each of these terms corresponds to
\begin{equation}
    \displaystyle \mathcal{Z}_1=\frac{k_0}{k_r} e^{L\left[\Delta \mu_F-\Delta \mu_r-\log(1-x_a)\right]}\int_0^{x_M} dx\,e^{-L\, D(x||x_a)},
    \label{eq5}
\end{equation}    
    \noindent and
\begin{equation}
    \displaystyle \mathcal{Z}_2=e^{L\left[\log m-\Delta \mu_r\right]}\int_{x_M}^1 dx\,e^{-L\, D(x||x_r)},
    \label{eq6}
\end{equation}    
    \noindent where $1-x_a=[1+e^{-a}(m-1)]^{-1}$, $x_r=(m-1)/m$, and $D(x||y)=x \log(x/y)+(1-x)\log[(1-x)/(1-y)]$ is the Kullback-Leibler divergence between two binary variables with probabilities $x$ and $y$. Note that $x_r$ is the expected fraction of errors when the copying procedure is performed at random, while $x_a$ is the solution obtained by Genthon et al. for the expected fraction of errors in the accurate regime as $L\to\infty$. From these terms, we can already see that we are recovering the phase diagram for large $L$ depicted in Fig.~2 in \cite{genthon2025nonequilibrium}. The first term $\mathcal{Z}_1$ would contribute to the partition function only when $\Delta \mu_F-\Delta \mu_r-\log(1-x_a)>0$, and the second term $\mathcal{Z}_1$ would contribute only when $\log m-\Delta \mu_r>0$. On the other hand, if $x_a<x_M<x_r$, both integrals become equal to 1 in the large $L$ limit and only the prefactors remain; otherwise, the minimum of the divergence is at the border $x=x_M$, and an extra term $\exp{[-L\, D(x_M||x_i)]}$ needs to be considered.

    If we are under the condition $x_a<x_M<x_r$, the partition function can be written as
\begin{equation}
    \displaystyle \mathcal{Z}=e^{L\left[\log m-\Delta \mu_r\right]}\left[\frac{k_0}{k_r}e^{L\left(\Delta \mu_F-\Delta \mu_F^*\right)}+1 \right],
    \label{eq7}
\end{equation}    
    \noindent where $\Delta \mu_F^*=\log m+\log(1-x_a)$ is the threshold in $\Delta \mu_F$ that determines which term of the partition function dominates. 

    Finally, $p(s|t)= \lambda_{xL}/\mathcal{Z}$, being $xL=q(s,t)$ the Hamming distance between the template and the copy. 

\begin{figure}
    \centering
    \includegraphics[width=1.\textwidth]{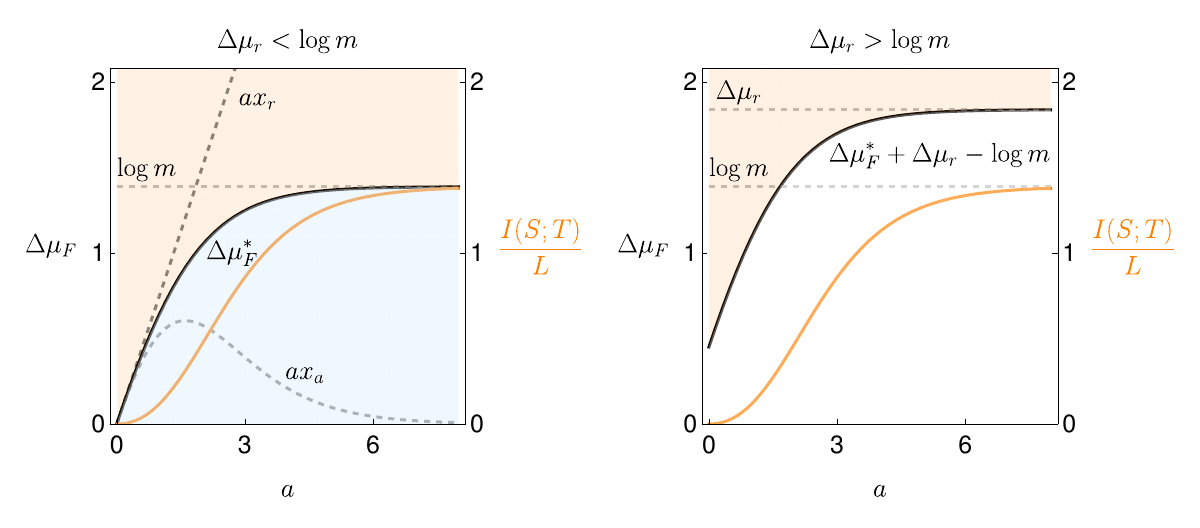}
    \caption{Phase diagram for mutual information. Left panel: Case $\Delta\mu_r<\log m$. Under this condition, the information is different from zero as long as $\Delta\mu_F>\Delta\mu_F^*$, corresponding to the orange colored region, while in the light-blue region random copies dominate making $I/L=0$. The amount of information per-monomer $I/L$ in this region only depends on $a$ (orange curve), increasing from zero to $\log m$ for large values of $a$. Right panel: Case $\Delta\mu_r>\log m$. Under this condition, the region where $I/L> 0$ becomes smaller, and it is given by $\Delta\mu_F>\Delta\mu_F^*+\Delta\mu_r-\log m$. In the white region, the population of copies vanishes. The dependency of $I/L$ with $a$ remains the same.}
    \label{fig1}
\end{figure}
    
    Due to the interchangeability of rows and columns of the conditional probability matrix, the replication process is equivalent to a symmetric channel. This fact implies that one input distribution $p(t)$ that maximizes the information is the uniform distribution, namely, $p(t)=m^{-L}$, which we use in the next section to calculate the mutual information. It is important to notice that this solution may not be unique, and other distributions could also reach the maximum. Moreover, in many contexts, it is biologically unrealistic that templates would actually exhibit such an input distribution\footnote{When coding information for a channel with capacity $C$ in blocks of length $n$, instead of using a uniform input, one could send $M=\exp(nC)\ll m^n$ messages at random with an exponentially (in $n$) small error at the decoding phase\cite{cover1999elements}. This solution may not be practical or plausible, but it highlights the multiplicity of ways to efficiently use such a channel.}.

\section{Mutual information in the ensemble}
    The mutual information characterizes how differently two stochastic variables behave in relation to the case where they are independent \cite{mackay2003information, cover1999elements}, which can be expressed as the Kullback-Leibler divergence between the joint distribution and the product of the marginals. The information can also be calculated as the reduction in entropy of one variable once we know the other, that is,
\begin{equation}
\begin{array}{rl}
     I(T;S)& =D\left[p(t,s)||p(t)p(s)\right]=H(S)-H(S|T)  \\\\
     & =L \log m- \sum_t p(t)\left[ -\sum_s p(s|t)\log p(s|t)\right]\\\\
     & =L \log m+\sum_s p(s|t_0)\log p(s|t_0).
\end{array}    
    \label{eq8}
\end{equation}
    
    Here, we have simplified this expression using two facts. First, due to the symmetry between all copies $s$, the marginal distribution of $S$ becomes uniform, that is, $H(S)=L \log m$. On the other hand, the conditional entropy has the same contribution for all $t$, so that $H(S|T)=H(S|T=t_0)$. Next, we replace the sum over $s$ with an integral over $x$, grouping all $\Omega_{x L}$ copies that are at the same distance $x(s,t_0)L$, and we use the fact that $p(s|t)= \lambda_{xL}/\mathcal{Z}$, 
\begin{equation}
    \displaystyle I(T;S)/L =\log m+\,\frac{1}{L} \int_0^1 dx\,\Omega_{xL}\frac{\lambda_{xL}}{\mathcal{Z}}\left(\log \lambda_{xL}-\log\mathcal{Z}\right).
    \label{eq9}
\end{equation}
    
    Here, we need Stirling's formula to show that $\log \Omega_{x L}\simeq L[x\log(m-1)+H(x)]$, where $H(x)$ is the entropy of a binary variable with probability $x$. Also, as we did before, we need to separate the integral in $x$ into two terms ($0<x<x_M$ and $x_M<x<1$). After some operations and after applying the Laplace method, we obtain
\begin{equation}
   \displaystyle I(T;S)/L =   
    \begin{cases}
      D(x_a||x_r) & \text{if $\Delta \mu_F>\Delta \mu_F^*$ $\land$ $\Delta \mu_F>\Delta \mu_r+\log(1-x_a)$}\\
      0 & \text{otherwise},
    \end{cases}    
    \label{eq10}
\end{equation}
    \noindent where $D(x_a||x_r)=\log m - x_a \log(m-1) - H(x_a)$. This relationship $D(x_a||x_r)$ between the information per-monomer and the fraction of errors is highly non-linear. When there are no errors $x_a=0$, that is, the limit of very high specificity, the information per-monomer goes to $\log m$, its maximum value. However, due to the divergence of the derivative $\partial_{x_a}D(x_a||x_r)$ in $x_a=0$, once there is a non-zero fraction of errors, the information per-monomer decreases substantially. As expected, the information vanishes when $x_a\to x_r$ ($a\to 0$).

\begin{figure}
    \centering
    \includegraphics[width=1.\textwidth]{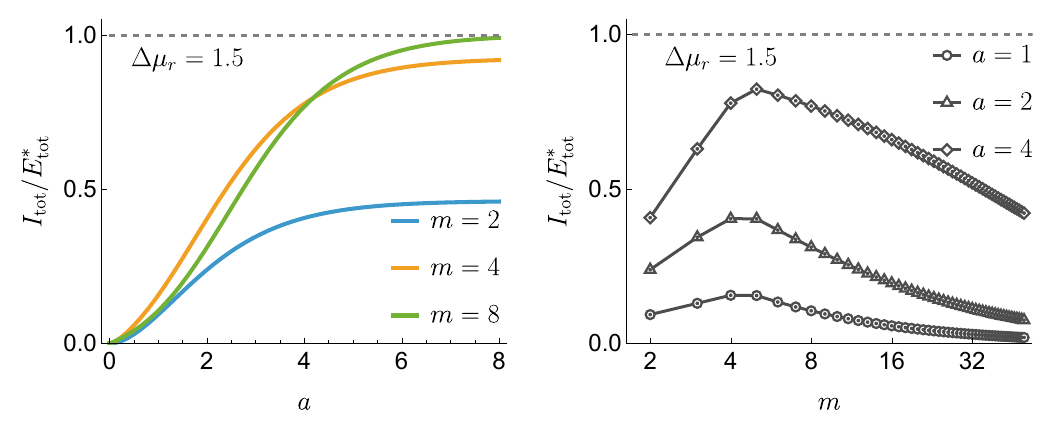}
    \caption{Information to energy cost ratio in a template copying ensemble. Left panel: Information to energy cost ratio as a function of template specificity $a$ for different number of monomers $m$, given a particular value of $\Delta\mu_r$. Right panel: Same ratio as a function of $m$ for different values of specificity $a$.}
    \label{fig2}
\end{figure}

    Naturally, given the calculations involved in the partition function, the phase diagram for mutual information per-monomer in the space $(a,\Delta\mu_F)$ in the large $L$ limit shares characteristics with the accurate-random regions \cite{genthon2025nonequilibrium} (see Fig.~\ref{fig1}, considering two cases for $\Delta\mu_r$). The information is different from zero when $\Delta\mu_F>\max(\log m,\Delta\mu_r)-\log[1+e^{-a}(m-1)]$, and the amount of information in such a region only depends on $a$, increasing with the specificity from zero to $\log m$. Let us notice that even though we are in an accurate region in the large $L$ limit, the amount of information may be well below its maximum, depending on the value of $a$ (orange curves in Fig.~\ref{fig1}).

    Having established when information transmission occurs, we now examine its efficiency. In general, obtaining copies with higher fidelity costs more energy. In this particular model, both the total information between templates and copies $I_{\text{tot}}=I(T;S)$ and the minimum total energy provided by the fuel to copy such sequences $E^*_{\text{tot}}=L\, \Delta\mu_F^{\text{(min)}}$ ---that is, the minimum energy to stay in the accurate regime--- grow linearly with the length of the chain. As we have seen from Eq.~(\ref{eq10}), the minimum free energy per-monomer provided by the fuel to reach the accurate region where $I(T;S)>0$ corresponds to $\Delta\mu_F^{\text{(min)}}=\max(\log m,\Delta\mu_r)-\log[1+e^{-a}(m-1)]$. As more information requires more energy, it becomes natural to question how much each bit of information costs in terms of energy provided by the fuel. For this purpose, we study the ratio of the total information between templates and copies to the minimum energy to copy such sequences, to quantify the efficiency of the replication system. This ratio corresponds to
\begin{equation}
    \displaystyle \frac{I_{\text{tot}}}{E_{\text{tot}}^*}=\frac{\log m -x_a \log(m-1)-H(x_a)}{\max(\log m,\Delta\mu_r)+\log(1-x_a)}.
    \label{eq11}
\end{equation}

    Assuming $x_a<(m-1)/m=x_r$, it follows that this ratio is bounded by 1 when $\log m>\Delta\mu_r$, and by $\log m/\Delta\mu_r$ otherwise. Given the fact that $I_{\text{tot}}/E_{\text{tot}}^*$ is a bounded quantity, it allows us to interpret it as a measure of efficiency. As a function of $a$, this ratio tends to the bound for large values of $a$ (see left panel in Fig.~\ref{fig2}). Now, when viewed as a function of the number of monomers $m$, this ratio of information-to-energy exhibits two behaviors. For low values of $m$, it increases until reaching a maximum at a certain value $m^*$, and then slowly decreases with $m$ (see right panel in Fig.~\ref{fig2}). For large values of $a$, it is evident that this peak occurs at $m^*\simeq \ceil*{e^{\Delta\mu_r}}$. The occurrence of the peak stems from the trade-off between the increased message space from a larger alphabet and the increasing energy cost to prevent random assembly.

    The existence of an optimal alphabet size has intriguing biological implications, particularly for DNA's four-base system ($m=4$). Our analysis suggests that for the information-to-energy ratio to be maximized at $m=4$, the effective assembly energy would need to be $\Delta\mu_r\approx \log 4\approx 1.4\,k_B T$. However, the actual effective assembly energy for DNA is substantially higher. Firstly, we need to consider the intrinsic energetic cost of forming the covalent phosphodiester bond of $+5.3$ kcal/mol ($8.6\,k_B T$) \cite{dickson2000determination}. Secondly, accounting for the contribution of individual base stacking to DNA stability \cite{abraham2023high} into the coarse-graining process results in a reduction of $-0.8$ to $-3.7\,k_B T$. Finally, the concentration values of the dNMP pool that determine the magnitude of $\mu_M$ in a living cell \cite{hill2012linear, bennett2009absolute} may fluctuates in the range $1\,\mu M$ to $100\,\mu M$, contributing with $+9.2$ to $+13.8\,k_B T$ to the assembly energy. All these terms total a value for the per-monomer assembly energy $\Delta\mu_r$ ranging from $+14$ to $+22\,k_B T$. This discrepancy suggests that the four-base alphabet is not optimized for information-per-unit-fuel efficiency. Instead, a high $\Delta\mu_r$ is critical for \emph{quenching} spontaneous assembly ($\Delta\mu_r>\log m$), ensuring that random polymers do not form in the absence of fuel-driven templates \cite{genthon2025nonequilibrium}. Thus, the choice of alphabet size reflects a trade-off where efficiency is sacrificed for the thermodynamic suppression of errors.


\begin{wrapfigure}{r}{0.5\textwidth}
  \begin{center}
    \vspace*{-15pt}
    \includegraphics[width=0.48\textwidth]{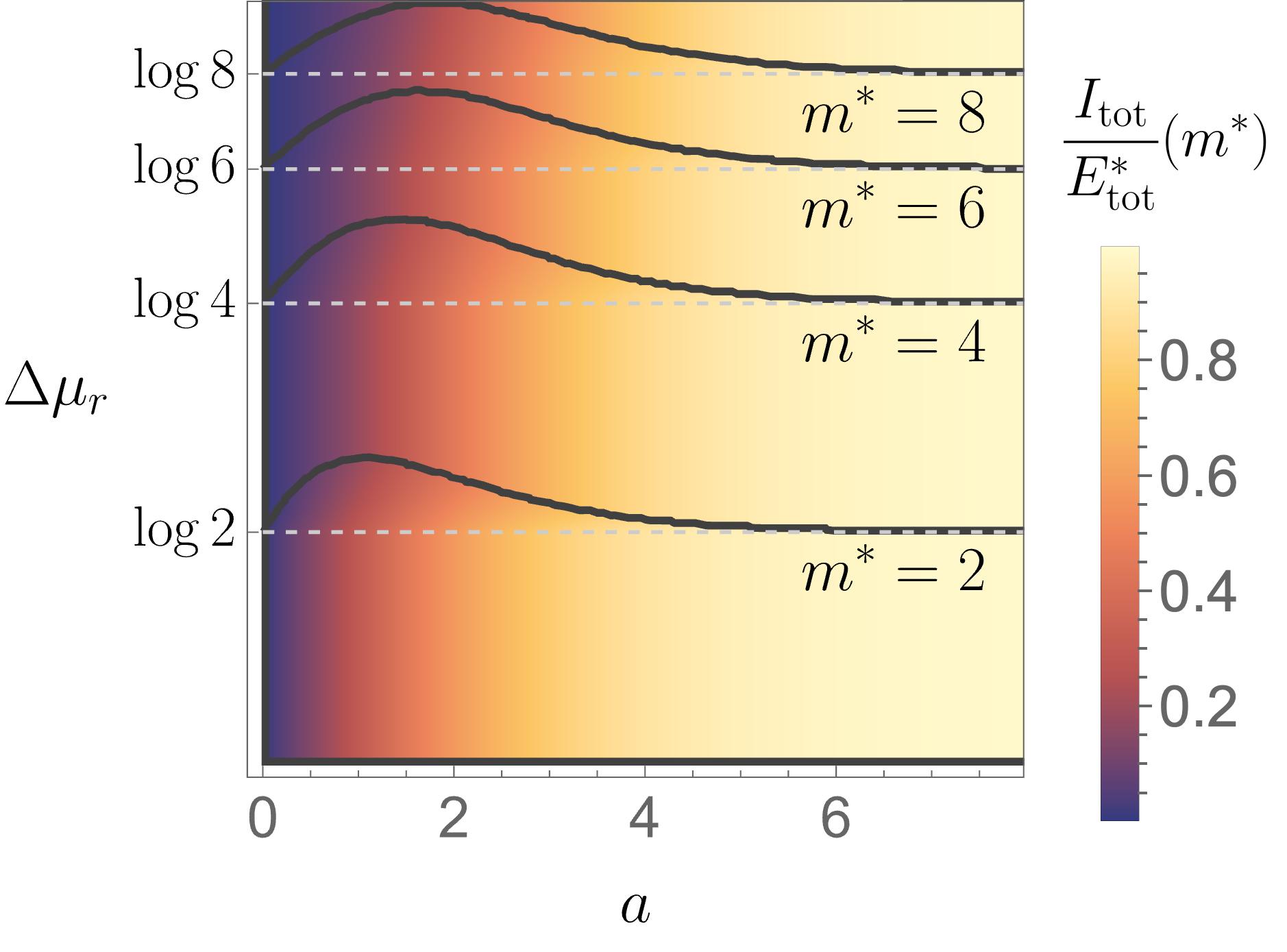}
    \vspace*{-15pt}
  \end{center}
  \caption{Optimal values of number of monomer types $m^*$ for the amount of information-to-energy cost (regions separated by black lines). Here only even values of $m^*$ are considered, and the colors in each region represent the maximum of the ratio $I_{\text{tot}}/E_{\text{tot}}^*$ for that value of $m^*$.}
  \vspace*{-70pt}
    \label{fig3}
\end{wrapfigure}

    As a final analysis for this section, we study what the optimal value for the number of bases $m^*$ in the space $(a,\Delta\mu_r)$ is, considering that the free energy of the fuel is at a minimum to achieve accurate copies and, in particular, in the case where the number of monomers needs to be \emph{even} to maintain pairing symmetry for the bases (see Fig.~\ref{fig3}). From Eq.~(\ref{eq11}) we see that for small values of $\Delta\mu_r$, the information to energy ratio decreases with $m$, and the optimum corresponds to two bases ($m^*=2$). Once $a$ is large enough, the optimum value corresponds to $m^*$ if $\log(m^*-2)<\Delta\mu_r<\log m^*$, and the ratio $I_{\text{tot}}/E_{\text{tot}}^*$ in that part of the phase space tends to $1$. 

\vspace{60pt}
\section{Extensions}
\label{sec4:extensions}

\subsection{Fundamental Limits in the Rate-Fidelity Space}
\label{sec4:shannon}

    Understanding the fundamental limits of information transmission in the template copying ensemble is essential for evaluating potential improvements through proofreading mechanisms \cite{bennett1979dissipation, murugan2012speed, rao2015thermodynamics}. Within the accurate regime, where the replication process produces copies with a fraction $x_a$ of errors determined by the specificity $a$, we can interpret the system as a communication channel and analyze achievable performance using rate-distortion theory \cite{mackay2003information}. Within this context, the rate $R$ represents the number of bits per monomer encoded in the template, rather than a chemical reaction time.

\begin{figure}
    \centering
    \includegraphics[width=1.\textwidth]{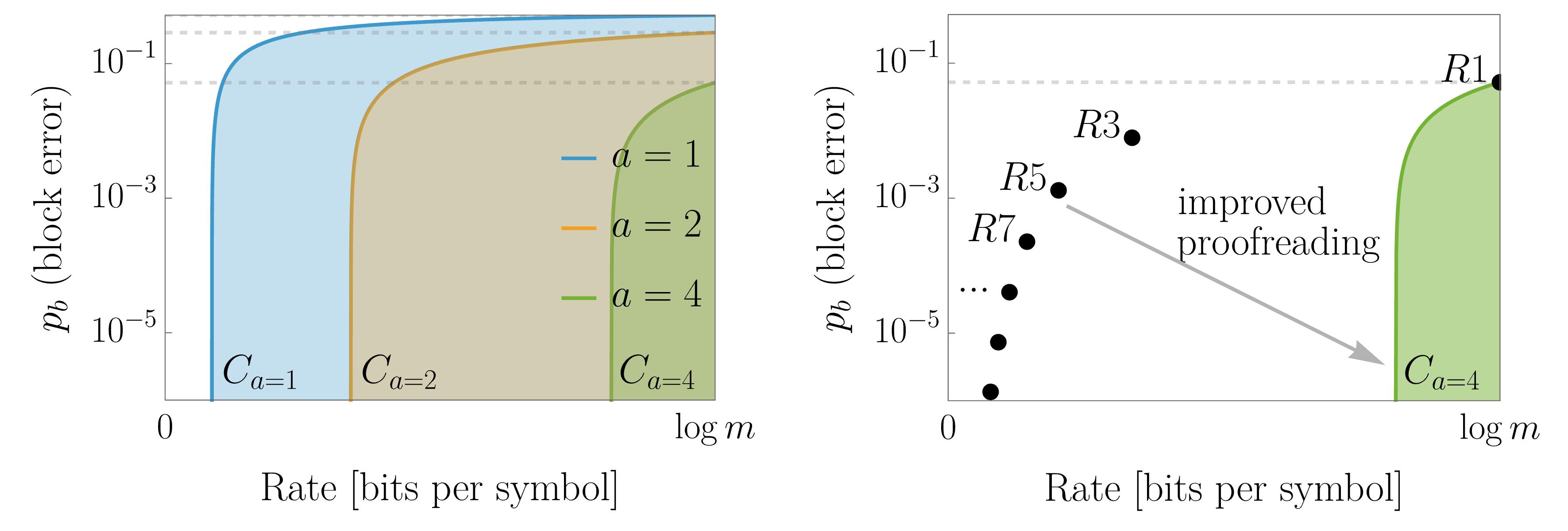}
    \caption{Limits in error reduction. Left panel: Accessible rate-error space. If we are using the system described by the template copying ensemble model to transmit information, we could improve the error $x_a=p_b(R=\log m)$ for individual symbols by reducing the rate. The shaded regions represent the inaccessible points in the space $(R,p_b)$ as given by the Shannon's bound. Here the capacity corresponds to $C_a=I_a/L$. Given an optimal code, we could obtain a performance of $p_b$ as small as desired using a length of chain increases by a factor $\log m/C_a$, or a lower speed of the overall process by a factor $C_a/\log m$. Right panel: Comparison of a repetition coding strategy in relation to Shannon's bound ($a=4$). A repetition code $Rn$ implies sending $n$ (odd) instances of each symbol and decode by a majority rule. The reduction in error by such an strategy \cite{mackay2003information} comes with a reduction in rate inversely proportional to $n$.}
    \label{fig4}
\end{figure}

    When operating at full speed ($R = \log m$), the channel produces copies with an error fraction $x_a$. However, by employing coding strategies ---essentially encoding information in blocks rather than copying symbol-by-symbol--- we can reduce the error rate at the cost of lowering the effective information transmission rate. Shannon's bound provides a fundamental limit on achievable rate-error combinations. For a desired block error probability $p_b$, only rates satisfying
\begin{equation}
    \displaystyle R< R(p_b)=\frac{C(x_a)\log m}{C(p_b)}=\frac{\log m\left[\log m - x_a \log(m-1)-H(x_a)\right]}{\log m - p_b \log(m-1)-H(p_b)},
    \label{eq12}
\end{equation}
    \noindent are achievable (see left panel of Fig.~\ref{fig4}). Here, the channel capacity $C(x_a)=\log m - x_a \log(m-1)-H(x_a)$ determines the maximum rate at which information can be reliably transmitted. Notably, arbitrarily small errors are theoretically achievable with the transmission speed reduced by a factor $C_a/\log m$.

    The distance between actual coding strategies and Shannon's bound measures their optimality. For instance, consider using $n$ copies $\{s_1,\dots,s_n\}$ to decode the template $t$ through majority-rule voting ---a repetition code $Rn$. While this strategy reduces errors, it does so inefficiently: the rate decreases as $1/n$ while remaining far from Shannon's bound (see the right panel of Fig.~\ref{fig4} for $a=4$). 
    Considering models where proofreading is implemented through an effective specificity $a_{\pm}$, as the one in \cite{genthon2025nonequilibrium}, the gap between repetition codes and the Shannon bound ---which is a theoretical maximum for the raw specificity before any proofreading-like processing--- illustrates the potential for more sophisticated error correction mechanisms.

    \textbf{Implications for proofreading models.} Future extensions of the template copying ensemble that incorporate proofreading mechanisms \cite{murugan2012speed, rao2015thermodynamics} should be evaluated against these fundamental limits. Any proofreading scheme induces a trade-off between copying speed and fidelity—the system must spend time (or equivalently, reduce throughput) to improve accuracy. Shannon's bound quantifies the best achievable trade-off: proofreading mechanisms that approach this bound are thermodynamically efficient, while those far from it waste resources on suboptimal error correction. For example, if a proofreading model achieves an error rate $p_b$ but requires reducing the effective copying rate to $R$, we can compare $R$ to Shannon's maximum $R(p_b)$ to assess optimality. In the terminology of information theory \cite{mackay2003information}, proofreading acts as a physical solution by modifying the kinetic characteristics of the copying channel. However, the Shannon bound of the underlying raw discrimination process provides a universal benchmark for evaluating the efficiency of such active noise-reduction mechanisms.

\subsection{Temperature Dependence and Thermodynamic Links}
    The template copying ensemble model as analyzed here assumes $k_B T=1$, effectively fixing the temperature. However, reintroducing explicit temperature dependence $\beta=(k_B T)^{-1}$ reveals how thermal fluctuations govern the information-accuracy phase transitions. Varying temperature is formally equivalent to rescaling the specificity $a$ relative to the energies $\Delta\mu_r$ and $\Delta\mu_F$, which modifies the phase boundaries (see Fig.~\ref{fig1}).

\begin{figure}
    \centering
    \includegraphics[width=1.\textwidth]{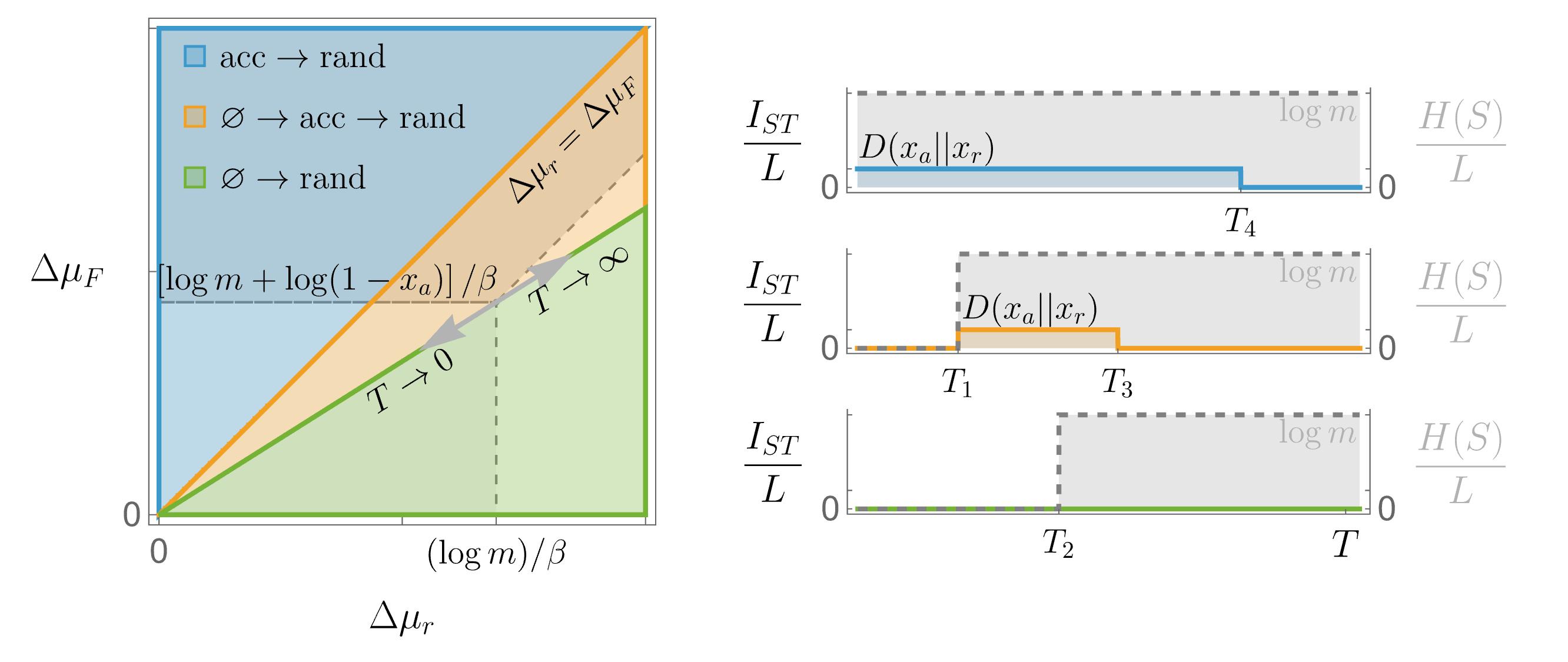}
    \caption{Transitions in information and entropy per-monomer when changing temperature. Left panel: Regions in $(\Delta\mu_r,\Delta\mu_F)$ that go through different transitions when changing temperature. In the blue region, the system goes from the accurate regime ($I/L=D(x_a||x_r)$, $H(S)/L=\log m$) to the random regime ($I/L=0$, $H(S)/L=\log m$) when increasing temperature. In the orange region, there are two transitions, first from no copies ($I/L=0$, $H(S)/L=0$) to an accurate regime, and then to a random regime. In the green region, there is only a single transition from no copies to a random regime. Right panel: Information and entropy per-monomer as a function of temperature. Corresponding transitions for a systems with the same value of $\Delta\mu_r$, and increasing value of $\Delta\mu_F$ from bottom to top. In both panels, $a=1.5$.}
    \label{fig5}
\end{figure}

    Figure~\ref{fig5} illustrates three distinct regimes depending on the values of $(\Delta\mu_r,\Delta\mu_F)$:
    \begin{itemize}
        \item Blue region: Systems transition from accurate copying ($I/L = D(x_a||x_r)$, $H(S)/L = \log m$) directly to random copying ($I/L = 0$, $H(S)/L = \log m$) as temperature increases. Here, thermal energy disrupts template specificity without eliminating copy production.
        \item Orange region: Two successive transitions occur ---first from no copies ($I/L = 0$, $H(S)/L = 0$) to accurate copying as fuel energy becomes effective, then to random copying at higher temperatures.
        \item Green region: Only a single transition from no copies to random copying, bypassing the accurate regime entirely.
    \end{itemize}

    Regarding the number of monomers $m^*$ that optimizes the ratio of information per unit of minimum fuel energy, the optimum now shifts with temperature as $m^* \approx e^{\beta \Delta\mu_r}$ ---that is, decreasing with higher temperatures to avoid the random assembly region.

    These temperature-induced transitions connect naturally to the work of Ouldridge and ten Wolde \cite{ouldridge2017fundamental}, who established that producing persistent copies with higher accuracy requires increasing thermodynamic work. In the template copying ensemble, once in the accurate region, fidelity is regulated solely by specificity $a$ (or equivalently, by temperature through $a/\beta$). A quantitative comparison requires care: Ouldridge and ten Wolde consider ensembles where copies from different templates can mix and potentially interact, whereas our framework treats each template's copy population independently. Nevertheless, both approaches should recover the fundamental principle that information preservation demands energy dissipation.

    Future work should make this connection explicit by (i) calculating entropy production rates as a function of $(a,\Delta\mu_r,\Delta\mu_F,\beta)$ to relate information $I(T;S)$ directly to dissipated work, and (ii) extending the ensemble model to allow copy-copy interactions, bringing it closer to the Ouldridge-ten Wolde framework.

\subsection{Proofreading Mechanisms in Ensemble Models}
    Several recent works have developed kinetic proofreading models that balance speed, accuracy, and energy dissipation \cite{bennett1979dissipation, murugan2012speed, rao2015thermodynamics}. Incorporating such mechanisms into the template copying ensemble framework would provide a natural testbed for information-theoretic analysis. Two approaches appear particularly promising:

    \textbf{Multi-stage verification.} Following Murugan et al. \cite{murugan2012speed}, we could extend the model to include intermediate checking states where partially assembled copies can be rejected before completion. Each checking stage consumes additional fuel (increasing $E_{\text{tot}}$) and time (decreasing rate $R$) but potentially reduces the error fraction $x_a$. The information-to-energy ratio $I_{\text{tot}}/E_{\text{tot}}$ would then depend on the number and stringency of checking stages, allowing optimization over proofreading architecture.
    
    \textbf{Kinetic discrimination.} Rao and Peliti \cite{rao2015thermodynamics} showed that thermodynamic efficiency in proofreading depends on how effectively the system exploits kinetic differences between correct and incorrect monomer incorporations. Within our ensemble framework, this corresponds to making the rates $k_a$ and $k_d$ in Eq.~(\ref{eq2}) depend not only on Hamming distance q but also on the specific sequence context. Such refinements would break the symmetry underlying Eq.~(\ref{eq1}) and require recalculating the partition function $\mathcal{Z}$, but would permit direct comparison between proofreading thermodynamics and information-theoretic limits.

    For both approaches, Shannon's bound provides a crucial benchmark: any proofreading mechanism achieving error $p_b$ at rate $R$ can be compared against the maximum theoretically achievable $R(p_b)$, revealing how much room for improvement remains.

\section{Discussion}

    We have recast the template copying ensemble of Genthon et al. \cite{genthon2025nonequilibrium} as an information transmission channel and calculated the mutual information $I(T;S)$ between templates and copies. Although we are still under the assumption that the generated copies are independent, this information-theoretic perspective reveals features not apparent in the original error-fraction analysis.

    First, the relationship between information and errors is highly nonlinear: $I/L = D(x_a||x_r) = \log m - x_a \log(m-1) - H(x_a)$. Because the derivative of $D(x_a||x_r)$ diverges at $x_a = 0$, even small error fractions cause substantial information loss. This matters because information—not error fraction—connects directly to entropy production \cite{andrieux2008nonequilibrium} and determines the fundamental limits of proofreading strategies through Shannon's bound.

    Second, and most strikingly, the existence of an optimal alphabet size determined by the assembly energy $\Delta\mu_r$ reveals that DNA's four-base system operates far from the theoretical optimum for information-to-energy efficiency. Given the estimated assembly energy of $\Delta\mu_r \gtrapprox 14\,k_B T$ \cite{dickson2000determination, abraham2023high, bennett2009absolute}, the information-to-energy ratio would favor a significantly larger alphabet size. The fact that biology utilizes a much smaller alphabet suggests that other constraints, such as the suppression of spontaneous random assembly and the necessity of high kinetic specificity, dominate the evolutionary landscape. Specifically, the high value of $\Delta\mu_r$ ensures that DNA sequences are naturally \emph{quenched} against non-templated synthesis, as $\Delta\mu_r$ significantly exceeds the $\log 4$ threshold required to prevent spontaneous polymerization \cite{landauer1961irreversibility, genthon2025nonequilibrium}. Consequently, the four-base system represents a regime where information-to-energy efficiency is traded for robust sequence preservation and active enzymatic control over replication.

    Third, Shannon's bound quantifies fundamental rate-fidelity trade-offs: achieving error $p_b$ requires reducing the transmission rate to $R < R(p_b)$. This provides a benchmark for evaluating proofreading mechanisms—biological systems like polymerase backtracking \cite{bennett1979dissipation, rao2015thermodynamics} can be assessed for thermodynamic efficiency by comparing their performance against this fundamental limit.

    Our analysis assumes uniform template distributions $p(t) = m^{-L}$, which lack biological realism but allow for analytical progress. Shannon's coding theorem \cite{mackay2003information, cover1999elements} indicates that near-optimal transmission remains possible with non-uniform inputs, suggesting robustness. The model also coarse-grains over molecular details (processivity, sequence-dependent kinetics, strand separation) and treats each template's copies independently, unlike Ouldridge and ten Wolde \cite{ouldridge2017fundamental} who allow mixing. Explicit entropy production calculations and copy-copy interactions remain important extensions.

    In summary, our study on the information balance of the template copying ensemble reveals that replication is not just about avoiding errors when copying a template but also about the efficient expenditure of fuel to preserve sequence information against the entropic pull of random assembly, with implications for understanding both natural replication fidelity and designing synthetic copying systems.

\vspace{20pt}
\textit{Acknowledgments}---The author would like to thank Arthur Genthon for helpful comments and engaging discussions. The author would also like to thank the Visitor Program of the MPI-PKS (Max Planck Institute for the Physics of Complex Systems) for its support during a visit to the institute in the spring of 2025. The author declares no conflict of interest.

\bibliographystyle{unsrt}
\bibliography{biblo.bib}

\end{document}